\documentclass[aps,prl,superscriptaddress,showpacs,floatfix]{revtex4}
\usepackage{graphicx}
\usepackage{amssymb}
\usepackage{amsmath}
\usepackage{color}

\newcommand{\unit}[1]{\ensuremath{\;\mathrm{#1}}}
\newcommand{\ket}[1]{\ensuremath{\left| #1 \right\rangle}}

\begin{document}

\title{Optimal excitation conditions for indistinguishable photons from quantum dots}

\author{Tobias Huber}
\author{Ana Predojevi\'{c}}
\author{Daniel F\"oger}
\affiliation{Institut f\"ur Experimentalphysik, Universit\"at  Innsbruck, Technikerstrasse 25, 6020 Innsbruck, Austria}
\author{Glenn Solomon}
\affiliation{Institut f\"ur Experimentalphysik, Universit\"at  Innsbruck, Technikerstrasse 25, 6020 Innsbruck, Austria}
\affiliation{Joint Quantum Institute, National Institute of Standards and Technology  and University of Maryland, Gaithersburg, MD 20849, USA }
\author{Gregor Weihs}
\affiliation{Institut f\"ur Experimentalphysik, Universit\"at  Innsbruck, Technikerstrasse 25, 6020 Innsbruck, Austria}

\begin{abstract}
In this letter, we present a detailed, all optical study of the influence of different excitation schemes on the indistinguishability of single photons from a single InAs quantum dot. For this study, we measure the Hong-Ou-Mandel interference of consecutive photons from the spontaneous emission of an InAs quantum dot state under various excitation schemes and different excitation conditions and give a comparison.
\end{abstract}

%chose pacs here%

\maketitle

\section{Introduction}
Quantum dots can emit single photons~\cite{Lounis00a, Michler00b, Straka14} as well as entangled photon pairs~\cite{Akopian06, Stevenson06, Dousse10, Jayakumar14}. 
Furthermore, the electron or hole spins in charged quantum dots could be used as local memories~\cite{Kroutvar04, DeGreve12}, which is a prerequisite to build quantum networks~\cite{Kimble08}. Those photons and photon pairs can be prepared deterministically through optical excitation~\cite{Jayakumar13, He13, Mueller14}, or the quantum dot can be pumped electrically~\cite{Benson00, Yuan02}. These photons are required for various quantum-enhanced tasks including linear optics quantum computation~\cite{Knill01}, quantum communication~\cite{Bennett95a}, and quantum sensing~\cite{Wolfgramm13}. Indistinguishable single photons on demand are essential in order to realize these technologies.

The first investigations of the indistinguishability of photons from quantum dots were performed by Santori et al. in 2002~\cite{Santori02b} using quasi-resonant excitation. Laurent et al. measured a similar result with quantum dots in a photonic crystal cavity in 2005~\cite{Laurent05} and Bennett et al.~\cite{Bennett05} investigated the influence of pump power when exciting the quantum dot quasi resonant  in 2005. Flagg et al. showed in 2012 that exciton recapture processes are responsible for single photon purity dynamics~\cite{Flagg12}. In 2008 Bennett et al. investigated the indistinguishability of photons from quantum dots in a diode structure under electrical pumping~\cite{Bennett08}. The first attempt at resonant pumping was performed in 2009 by Ates et al.~\cite{Ates09}, and the value for the indistinguishability was improved with resonant pumping in 2013 by He et al.~\cite{He13}. In 2013 Gazzano et al. showed that good indistinguishability can be achieved in a high brightness source~\cite{Gazzano13}.  Stevenson et al.~\cite{Stevenson12} showed that a diode structure can produce indistinguishable and entangled photons with electrical pumping. The first to show indistinguishable photons from the biexciton state were M\"uller et al. in 2014~\cite{Mueller14}. Also the indistinguishability of photons from two quantum dots~\cite{Patel10, Flagg10}, from a quantum dot photon and a photon generated by a SPDC source~\cite{Polyakov11} and between a quantum dot photon and a laser~\cite{Bennett09} has been studied. Thus, with improved experimental techniques and device engineering over the years, there has been steady improvement in the quantum properties and brightness from quantum dot photons.

In addition, theoretical studies of two photon interference~\cite{Legero03} and experimental realizations with photons created by atom-cavity systems~\cite{Legero04} showed that apart from the spatial and the polarization overlap of the photons the quality of the wavepackets also plays an important role. Here, especially important are the dephasing and the jitter of the arrival time of the wave packets.
To better understand these effects in a quantum dot system we have made a comparison study.
We measured the interference of the fields from both exciton and biexciton photons from an InAs quantum dot under different excitation conditions in a Hong-Ou-Mandel (HOM) interferometer.
In particular, we measured the indistinguishability of exciton photons with a varying excitation pulse length and a varying detuning of the excitation with respect to a quasi resonance. Additionally, we performed an indistinguishability measurement in resonant excitation of the biexciton. In this process the exciton state is populated from the radiative biexciton decay, resulting in an additional time-jitter compared to quasi resonant excitation of the exciton. The quasi resonant excitation is not time-jitter free since phonon relaxation is needed to populate the exciton state, but the time-jitter is much smaller than the biexciton decay, because it is non radiative. The phonon relaxation time is in the order of a few ps~\cite{Heitz97}, whereas the radiative lifetime of the biexciton is 370\unit{ps}.

For the investigations of the indistinguishability of the biexciton photon, we performed two different excitation schemes - resonant excitation and above band excitation. For resonant excitation of the biexciton state we use a two-photon excitation exploiting a virtual level~\cite{Jayakumar13}. For comparison, we performed a HOM interference measurement on the biexciton recombination photon with the quantum dot excited above-band, where phonon transitions are needed to relax to the biexciton state. 
This pump scheme introduces two mechanisms that reduce the indistinguishability: first, the time uncertainty of the relaxation process, which causes timing jitter~\cite{Santori04, Flagg12,Kiraz05} and second, charge noise induced by charge carriers in the vicinity of the quantum dot, which shifts the quantum dot resonances through varying potentials.~\cite{Bennett05}
\\

\section{Methods}
{\em Theory:}
If two indistinguishable photons simultaneously impinge on a beamsplitter, they coalesce and leave the beamsplitter through the same output port. This effect was first described by Hong, Ou, and Mandel~\cite{Hong87a} and is therefore referred to as Hong-Ou-Mandel (HOM) interference.
Let us consider a beamsplitter with two input modes (enumerated as 1 and 2) and two output modes (enumerated as 3 and 4). The four propagating field components  $\hat{E}_{1-4}^{(+)}$ are related by the transformation
\begin{equation}\label{eq:fields}
\begin{aligned}
\hat{E}_3^{(+)}=\sqrt{T}\hat{E}_1^{(+)}+i\sqrt{R}\hat{E}_2^{(+)}\\
\hat{E}_4^{(+)}=\sqrt{T}\hat{E}_2^{(+)}+i\sqrt{R}\hat{E}_1^{(+)},
\end{aligned}
\end{equation}
where R and T are the reflection and transmission coefficient of the beamsplitter, respectively.
If one considers a system with pure dephasing (which leads to an exponential coherence time, $T_2$) and an exponential lifetime of the excited state $T_1$, the probability to find two photons at opposite outputs is
\begin{equation}
\label{eq:P}
P(\tau)=\frac{1}{2}-\frac{1}{2}\frac{2RT}{1-2RT}\frac{T_2 }{2T_1-T_2}\left[e^{-2|\tau|/T_1}-\frac{T_2}{2T_1}e^{-4|\tau|/T_2}\right],
\end{equation}
where $\tau$ is the difference of the arrival time of the photons on the beamsplitter~\cite{Bylander02}. Timing jitter, which is an additional time uncertainty in the excitation process, produces an effective decoherence with a Gaussian shape instead of an exponential one. This contribution cannot clearly be distinguished in our data and is therefore not considered separately. Spectral diffusion, which is the wandering of the emission line due to interaction of the quantum dot with its environment, happens at much longer timescales and does not need to be considered in our case.
To simplify the above formula, let us consider a perfect beamsplitter with $R=T=0.5$. In this case, the factor $\frac{2RT}{1-2RT}$ equals one and the minimal observable value of the two photon coincidence probability is given by $P(0)=(1-\frac{T_2}{2T_1})/2$.
Here, we can see that for perfect indistinguishability $P(0)=0$ occurs when $T_2=2T_1$.
\\

\begin{figure}[ht]
\begin{centering}
\includegraphics[width=\columnwidth]{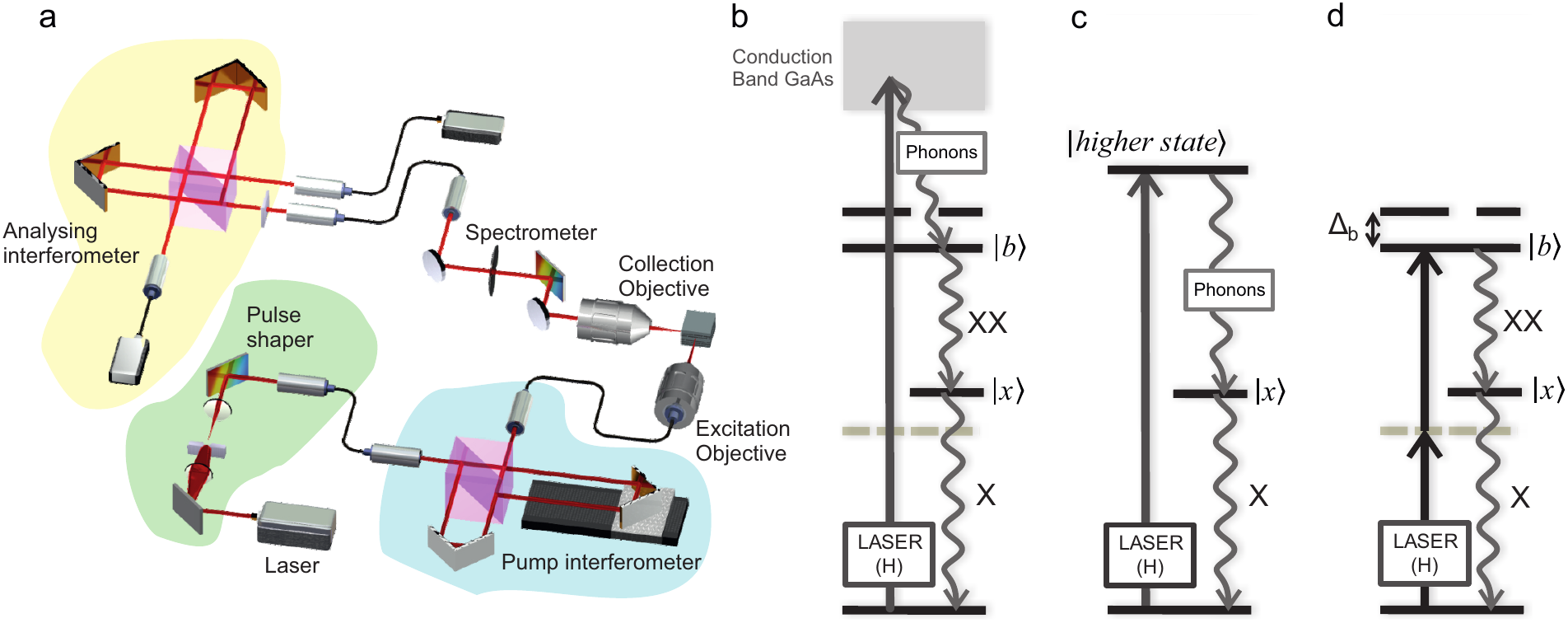}
\par\end{centering}
\caption{\label{fig:setup}\textbf{a)}~Experimental setup. \textbf{b-d)}~Excitation schemes. \textbf{a)}~A pulse stretcher is used to vary the pulse length of a 2.5\unit{ps} Ti:Sapphire laser. Afterwards the light passes through the pump interferometer which creates two consecutive laser pulses. These two laser pulses are used to excite the quantum dot, which emits two successive photons. The exciton and biexciton photons are separated in wavelength and are sent through an analyzing interferometer which has the same imbalance as the pump interferometer. \textbf{b)}~Above-band excitation: carriers are excited in the conduction band of the surrounding material and relax phonon assisted to the quantum dot states.  \textbf{c)}~Quasi resonant excitation scheme, where a higher, short lived bound state of the quantum dot is excited resonantly with the laser and decays to the exciton state \ket{x} and \textbf{d)}~two-photon resonant excitation of the biexciton state.}
\end{figure}

{\em Experimental Setup:} Our sample contains a layer of self-assembled InAs quantum dots grown by molecular-beam epitaxy, embedded in a $4\mathrm{\lambda}$ planar distributed Bragg reflector (DBR) cavity. The cavity consists of 15.5 lower and 10 upper layers of AlAs and GaAs to form the thwo DBRs. This sample was also used for previous work~\cite{Jayakumar13, Jayakumar14}. The sample is kept inside a liquid helium flow cryostat at 5\unit{K} and is excited orthogonally to the collection direction exploiting the lateral mode of the DBR cavity for wave-guiding~\cite{Muller07}. The quantum dot emission is collected from the top using an objective (NA=0.68), forwarded through a spectrometer, and sent into single mode fibers. A schematic of the experimental setup is shown in Figure~\ref{fig:setup}a. Light derived from a pulsed Ti:Sapphire laser with a repetition rate of $81\unit{MHz}$ is guided through a pulse stretcher, which consists of two gratings and a slit in 4-f configuration. The slit width can be varied to create pulses of different lengths (2.5\unit{ps} to 8\unit{ps}). These pulses travel through the pump interferometer, which is an imbalanced Michelson interferometer with a variable path-length difference from 2 to 4\unit{ns}. Thereby two successive pump pulses with variable time delay are created. The emission from the quantum dot is sent to the HOM interferometer, which is an imbalanced Michelson interferometer with 3\unit{ns} path length difference. If the early and late photons travel the long and short arms of the HOM interferometer, they arrive at the beamsplitter of the interferometer at the same time and will interfere. Changing the polarization or delay of one of the photons introduces distinguishability. The two outputs of the HOM interferometer are coupled to single mode fibers. The photons are then detected with avalanche photo diodes. In front of the HOM interferometer the emission can be sent to an intensity modulator which is a $\mathrm{LiNbO_3}$ Mach-Zehnder modulator with a bandwith of 20\unit{GHz}. The intensity modulator is optically synchronized to the laser and has a pulse width of 200\unit{ps} with  Gaussian pulse shape.

The three different excitation schemes used in our study are above-band excitation (excitation wavelength $\lambda_{exc,ab}=800\unit{nm}$), quasi resonant excitation  (excitation wavelength $\lambda_{exc,qr}=879.15\unit{nm}$) and two-photon resonant excitation to the biexciton state (excitation wavelength $\lambda_{exc,2p}=918.5\unit{nm}$, Figure~\ref{fig:setup}~b-d). In  above-band excitation a laser with an energy higher than the GaAs band-gap is used. This laser creates charge carriers in the material surrounding the quantum dot, which cascade through various pathways to the inhomogeneously broadened quantum dot distribution, where they can randomly fill quantum dots states (see Figure~\ref{fig:setup}~b). Exciton \ket{x}, biexciton \ket{b} and charged states can be created probabilistically with above-band excitation.
 In the method of quasi resonant excitation, 
a laser resonantly excites a higher excited state of the quantum dot. The higher excited quantum dot state can then decay to the exciton state which then decays emitting a photon of interest (see Figure~\ref{fig:setup}~c). In the two-photon resonant scheme the biexciton state \ket{b} is resonantly excited by exploiting a two-photon resonance via a virtual level. This virtual transition is at half the biexciton energy and it is not resonant with the exciton. The exciton and biexciton are anharmonic due to different Coulomb binding~\cite{Jayakumar13} (see Figure~\ref{fig:setup}~d).

\section{Results}

\begin{figure}[ht]
\begin{centering}
\includegraphics[width=0.9\columnwidth]{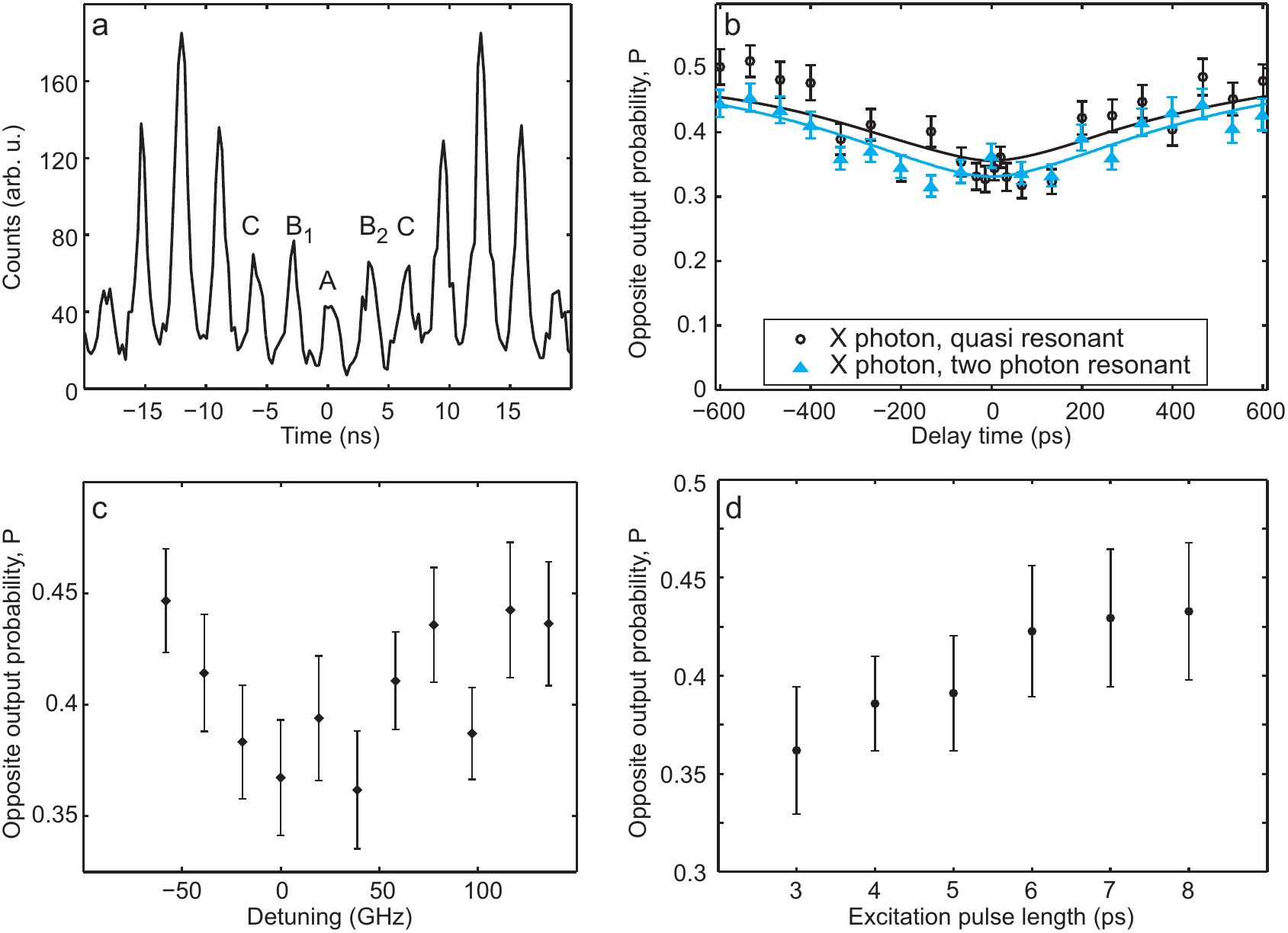}
\par\end{centering}
\caption{\label{fig:p-shell}HOM interference for the exciton emission obtained under quasi resonant excitation. \textbf{a)}~single point measurement, which is a histogram of a start-stop measurement between the two outputs of the analyzing interferometer. The origin of the multiple peaks is explained in the text. \textbf{b)}~Each single point is a measurement as shown in a) with a varying time delay between the two impinging photons. The time delay was introduced by changing the longer arm position in the pump interferometer. The black circles are the results for quasi resonant excitation and the blue triangles are the results for the exciton when exciting the biexciton two-photon resonantly. \textbf{c)}~A $3\unit{ps}$ laser pulse was tuned over a quasi resonance while all the other experimental parameters like excitation power and cryostat temperature were kept constant. \textbf{d)}~Exciting the quantum dot with different pulse lengths via a quasi resonance influences their indistinguishability. For our system parameters shorter pulses are advantageous.}
\end{figure}

The HOM measurement results for the exciton in quasi resonant excitation are shown in Figure~\ref{fig:p-shell}. Figure~\ref{fig:p-shell}~a shows a histogram of the arrival time differences of the photons on the two detectors. 
The five peaks (A-C) have their origin in three different types of coincidence events. The quantum dot is excited with two pulses, separated by $3\unit{ns}$, created by the pump interferometer. These two photons enter the HOM interferometer. If the first photon follows the short arm and the second photon follows the long arm, the two peaks C in Figure~\ref{fig:p-shell}~a at $\pm6\unit{ns}$ are recorded (peaks C from all laser pulses are overlapped because the laser repetition is $12\unit{ns}$). For peaks $\mathrm{B_1}$ and $\mathrm{B_2}$  at $\pm3\unit{ns}$, both photons follow the same arm. For peak A the first photon follows the long arm and the second photon follows the short arm. Therefore these two photons will arrive at the beamsplitter simultaneously and can interfere.

The peaks A-C are created from two single photons which results in a ratio between A:B:C of 2:2:1 if they do not interfere and 0:2:1 if they perfectly interfere. The peaks from 6 to 20\unit{ns} are created from photons that have no correlations and therefore obey Poissonian statistics. The ratio A:B:C for a light source with Poissonian statistics is 6:4:1. This is important because it is not possible in our setup to keep the count rate perfectly constant over all the performed measurements. To achieve comparability of the results for different detector count levels, the area of peak A is normalized with the area of the peaks $\mathrm{B_{1,2}}$ using the formula $P(\tau)=\frac{A}{B_{1}+B_{2}}$, where $\tau$ describes the time delay between the photons for different measurement settings. With this definition P is equal to the probability to detect two photons at the different outputs of a beamsplitter.~\cite{Santori02b}  

Each point in Figures~\ref{fig:p-shell}~b-d is the result of a measurement in Figure~\ref{fig:p-shell}~a with a varying time delay between the two impinging photons. The time delay was introduced by changing the longer arm position in the pump interferometer. The indistinguishability for photons from an exciton decay is shown in Figure~\ref{fig:p-shell}~b, as a function of photon arrival time delay. The data is fitted with the model from Equation~\ref{eq:P}. For the black circles, which represent the measured data points for quasi resonantly excited excitons, the fit function gives a lifetime of $T_1=800(40)\unit{ps}$ and a coherence time of $T_2=450(35)\unit{ps}$. The fitted lifetime is consistent with a standard lifetime measurement (directly measuring the exciton decay) which results in $T_1=820(15)\unit{ps}.$
The coherence time is different from the single photon coherence time ($g^{(1)}$) which resulted in $T_2=95(5)\unit{ps}$, measured with the same pump conditions. Such a difference is expected due to the different time scales on which the two coherence times are considered, as other authors have reported~\cite{Santori02b, Bennett05, Mueller14, He13}
The blue triangles in Figure~\ref{fig:p-shell}~b represent the measured data points for exciton photons when exciting the quantum dot resonantly to the biexciton via a two-photon resonance. The fit result is $T_1=800(90)\unit{ps}$ and $T_2=540(50)\unit{ps}$, producing slightly better and statistically significantly improved coherence times.

Figure~\ref{fig:p-shell}~c shows the influence of the detuning of the excitation laser frequency from the quasi-resonance condition. We performed HOM-measurements for zero photon arrival time delay for frequency detunings from -60 to 140\unit{GHz} in 20\unit{GHz} steps. The excitation power was kept constant. To perform this measurement, we set a fixed slit width in the pulse stretcher to produce 3\unit{ps} pulses and moved the slit position to introduce detuning. The amount of detuning was measured using a spectrometer. Due to the detuning the quasi resonant pump process gets less efficient and incoherent pump processes can contribute more to the measured signal. This leads to a decreased indistinguishability of the emitted photons. 
Figure~\ref{fig:p-shell}~d shows the influence of the excitation pulse length on the indistinguishability of the emitted photons. 
Here, the slit width of the pulse stretcher was modified while the central wavelength was kept at zero detuning. The pulse duration was measured with an autocorrelator and the given values represent the deconvoluted pulse envelope time. The excitation power was kept constant. The measurement result shows that longer excitation pulses  reduce the indistinguishability of the emitted photons. This could have several reasons. Obviously, an additional time uncertainty caused by the longer pulse is added in the excitation process. Also multiple excitations could occur, since the probability for the exciton state to decay, while the pump pulse is still present, increases. 
Thus, for our system, these effects are more significant than the disadvantages of a shorter pulse; for instance, large bandwidth and high peak intensity which leads to off-resonant generation of excitons.

\begin{figure}[ht]
\begin{centering}
\includegraphics[width=\columnwidth]{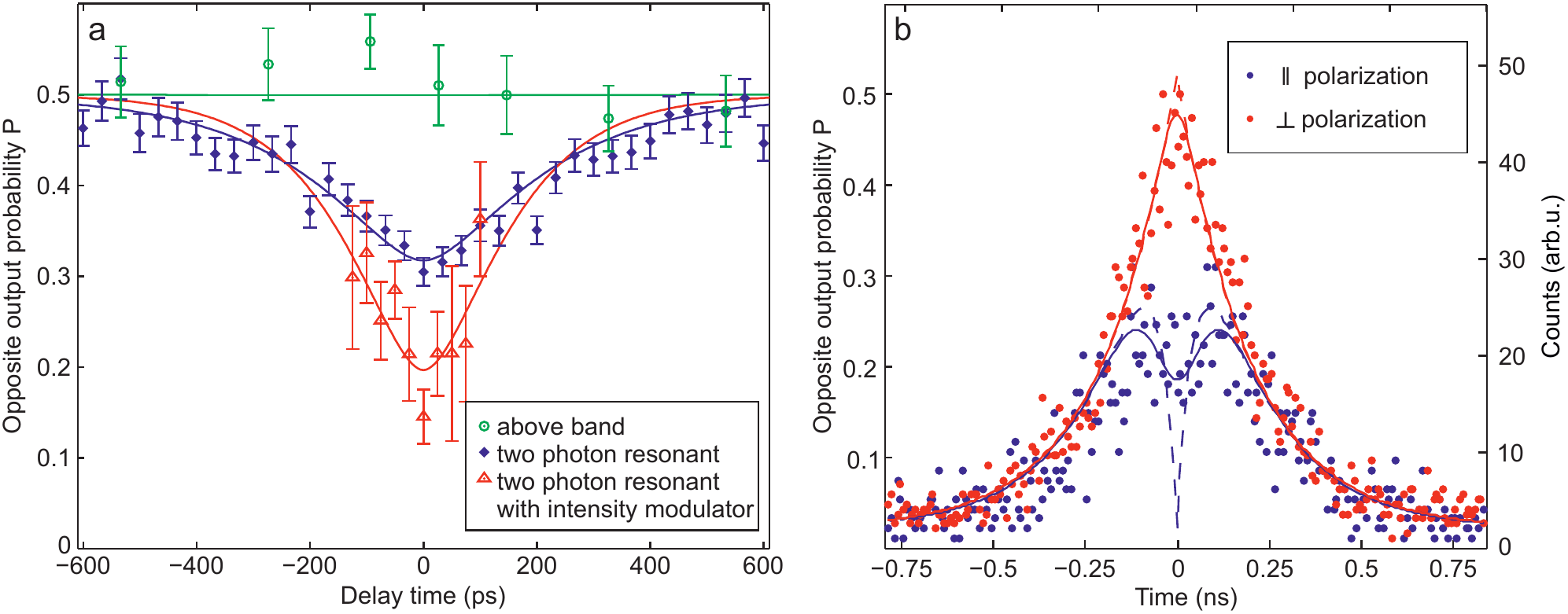}
\par\end{centering}
\caption{\label{fig:xx-hom} \textbf{a)}~HOM interference result of the biexciton recombination photon (XX) with another XX photon created from two successive pump pulses under resonant two photon excitation (2p-resonant). The blue filled circles show the opposite output probability for XX photons as emitted and the red triangles show XX photons after being cut to 200\unit{ps} wave packets by an intensity modulator. The solid lines represent fits to the data where lifetime and coherence time are the two fit parameters. The coherence time is fitted to both data sets simultaneously, because we do not change it with the intensity modulator. The green open circles show the same measurement with an above-band pump, where no indistinguishability was measured.
\textbf{b)}~Zoom in on peak A (see Figure~\ref{fig:p-shell}a) for the biexciton photon for zero time delay without intensity modulator with high timing resolution detectors fiber-pigtailed single-photon detection modules from Micro Photon Devices with a resolution of $35\unit{ps}$). The time axis here is the time between two detection events as in Figure~\ref{fig:p-shell}a and not the delay time between the two photons. The blue data points represent a measurement where the impinging photons have parallel polarization and will interfere on the beamsplitter, whereas the red data points represent a measurement with orthogonally polarized photons, which do not interfere because they are distinguishable in their polarization. The dashed lines are theoretical curves for infinite detector timing resolution with the given coherence time and lifetime. The solid lines are a convolution of the theoretical curve with the detector resolution of $35\unit{ps}$. }
\end{figure}

In Figure~\ref{fig:xx-hom} the result of the Hong-Ou-Mandel measurement for the biexciton photons is depicted. 
Here, we performed the measurement with a two-photon resonant excitation scheme on the biexciton state. For comparison we also performed a measurement using above-band excitation. The comparison with a quasi resonantly excited biexciton state is not possible because the quasi resonant excitation does not decay through the biexciton state.
The result for above-band excitation is depicted as green open circles. Under these excitation conditions the two photons appear to be completely distinguishable, although the (one photon) coherence time is about the same as for the resonant excitation case (140(10)\unit{ps} and 211(5)\unit{ps}, respectively). Timing jitter could be responsible for this behavior because it is only present in the above-band excitation case. The blue filled circles represent measurements where the quantum dot was excited directly to the biexciton, exploiting a two photon resonant excitation. To further improve the HOM interference, we set up an intensity modulator, which is an interferometer with an electro-optic phase modulator to create constructive or destructive interference on the output beamsplitter. With this pre-selection technique the shape of the quantum dot photon wave packet can be modified~\cite{Kolchin08, Specht09, Ates13}. This shaping shortens the photon wave packet, mimicking a shorter lifetime and thereby improves the ratio of coherence time to lifetime. The fitted lifetime for the blue filled circles (two photon excitation) is $T_1=375(30)\unit{ps}$ which agrees very well with the externally measured value for the lifetime of $T_1=370(10)\unit{ps}$ and the lifetime for the red triangles (two photon excitation and shortening by intensity modulator) is $T_{1\mathrm{,shaped}}=220(15)\unit{ps}$. The two curves were fitted simultaneously, because the coherence time is not influenced by the shaping procedure and the common value is $T_2=270(15)\unit{ps}$. Employing the intensity modulator improves the $\mathrm{T_1}:\mathrm{T_2}$ ratio, but does not yet allow for $2\mathrm{T_1}=\mathrm{T_2}$, as shown in Figure~\ref{fig:xx-hom}~a.

\section{Discussion}

We show that our initial result of 0.39(2) for the the indistinguishability of the biexciton photon can be improved  to 0.71(3) by applying time filtering.
For comparison, the result for the biexciton indistinguishability from M\"uller et al.~\cite{Mueller14} is 0.58(4) (uncorrected); however the lifetimes in~\cite{Mueller14} are shorter and the coherence lengths are approximately the same as reported here.

We also want to compare our result to the best reported values for resonant excitation of an exciton state, which are 0.82(10) from Gazzano et al.~\cite{Gazzano13} and 0.91(2) from He et al.~\cite{He13}. Both of these works used a sample structure with micro-pillar cavities to shorten the lifetime of the exciton state and thereby drastically improving the $\mathrm{T_1}:\mathrm{T_2}$ ratio. All of the given values are uncorrected for two reasons: first, one cannot distinguish a background event from a detected event, because the detector cannot discriminate between a thermal electron or a photon. Therefore, we think that subtracting the background, although its amount is known, is not a valid operation. Second, if one wants to build a scalable network with the given resources, the real, measured value is of interest and not a corrected one, because it is not possible to correct for the error while the photon is propagating through a network. All errors given by finite mode-overlap or remaining multi-photon contribution will add up and decrease the final process fidelity. We can speculate that if we would apply a micro-pillar cavity with a Purcell-factor of $\mathrm{F_P}=3$, the biexciton photon would show an indistinguishability of one. A micro-pillar cavity with such a Purcell-factor is rather standard, but the suppression of the scattered excitation laser would be the challenge in such a case.

\section{Conclusion}
We compared different excitation schemes and conclude that the pump scheme as well as the exact pumping conditions influence the indistinguishability of single photons emitted from a single semiconductor quantum dot. As reported before~\cite{He13, Mueller14}, resonant excitation leads to the highest quality photons. Additionally, we showed the amount of improvement in indistinguishability of the biexciton photon between resonant two-photon excitation and above-band excitation is 39(2)\unit{\%}. Further we showed that time filtering using an intensity modulator can improve the indistinguishability by improving the ratio of the lifetime to the coherence time $\frac{T_2}{2T_1}$. In our case this improved the indistinguishability to 71(3)\unit{\%}. This leads to the conclusion that for improving the quality of single photons to perform linear optics quantum computation, quantum communication, and quantum sensing, quantum dots must be pumped resonantly. The quality of the photons can be enhanced to the desired level if their lifetime is shortened, either by elaborate cavity designs~\cite{Vuckovic03, Dousse10, Gazzano13, He13} or by applying time filtering. 
\\
\begin{acknowledgments}
\section{Acknowledgments}
The authors thank the funding agencies. This work was funded by NSERC, the European Research Council (project EnSeNa, no. 275531). T. H. thanks the Austrian Academy of Sciences (\"OAW) for receiving a DOC Fellowship. A. P. would like to thank Austrian Science Fund (FWF) for support provided through Elise Richter Fellowship V-375. G.S.S. acknowledges partial support from the NSF PFC@JQI, and from Fulbright Austria – Austrian American Educational Commission through the Fulbright-University of Innsbruck Visiting Scholar program.
\end{acknowledgments}

\bibliographystyle{IEEEtran}
%\bibliography{C:/Documents/qobib}


\end{document}